\documentclass[11pt,twoside]{article}
\usepackage{./asp2014}
\usepackage{natbib}

\aspSuppressVolSlug
\resetcounters

\begin{document}

\title{Hubble Catalog of Variables}
\author{M. Yang$^1$ A. Z. Bonanos,$^1$, P. Gavras$^1$, K. Sokolovsky$^{1,2,3}$, D. Hatzidimitriou$^{1,4}$, M. I. Moretti$^5$, A. Karampelas$^1$, I. Bellas-Velidis$^1$, Z. Spetsieri$^{1,4}$, E. Pouliasis$^{1,4}$, I. Georgantopoulos$^1$, V. Charmandaris$^1$, K. Tsinganos$^1$, N. Laskaris$^6$, G. Kakaletris$^6$, A. Nota$^{7,8}$, D. Lennon$^9$, C. Arviset$^9$, B. Whitmore$^7$, T. Budavari$^{10}$, R. Downes$^7$, S. Lubow$^7$, A. Rest$^7$, L. Strolger$^7$, and R. White$^7$}

\affil{$^1$IAASARS, National Observatory of Athens, Penteli 15236, Greece; \email{myang@noa.gr}; \email{bonanos@noa.gr}}
\affil{$^2$Astro Space Center of Lebedev Physics Institute, Moscow 117997, Russia}
\affil{$^3$Sternberg Astronomical Institute, Moscow State University, Moscow 119992, Russia}
\affil{$^4$Department of Physics, National and Kapodistrian University of Athens, Ilissia 15771, Greece}
\affil{$^5$INAF-Osservatorio Astronomico di Capodimonte, Napoli 80131, Italy}
\affil{$^6$Athena Research and Innovation Center, Maroussi 15125, Greece}
\affil{$^7$Space Telescope Science Institute, Baltimore, MD 21218, USA}
\affil{$^8$European Space Agency, Research and Scientific Support Department, Baltimore, MD 21218, USA}
\affil{$^9$European Space Astronomy Centre, Madrid 28692, Spain}
\affil{$^{10}$The Johns Hopkins University, Baltimore, MD 21218, USA}

\begin{abstract}
The Hubble Catalog of Variables (HCV) project aims to identify the variable sources in the Hubble Source Catalog (HSC), which includes about 92 million objects with over 300 million measurements detected by the WFPC2, ACS and WFC3 cameras on board of the Hubble Space Telescope (HST), by using an automated pipeline containing a set of detection and validation algorithms. All the HSC sources with more than a predefined number of measurements in a single filter/instrument combination are pre-processed to correct systematic effect and to remove the bad measurements. The corrected data are used to compute a number of variability indexes to determine the variability status of each source. The final variable source catalog will contain variables stars, active galactic nuclei (AGNs), supernovae (SNs) or even new types of variables, reaching an unprecedented depth ($V\leq27~mag$). At the end of the project, the first release of the HCV will be available at the Mikulski Archive for Space Telescopes (MAST) and the ESA Hubble Science Archives. The HCV pipeline will be deployed at the Space Telescope Science Institute (STScI) so that an updated HCV may be generated following future releases of HSC.
\end{abstract}

\section{Introduction}

As the most influential astronomical mission, the Hubble Space Telescope (HST) has produced an unprecedented archive of deep, high-resolution images covering the near ultraviolet, optical and near infrared bands in its continuous operation over the past 26 years. The Hubble Source Catalog (HSC) has been built as a single master catalog based on this wealth of information by adopting a uniform reduction of the majority of publicly available images obtained with the WFPC2, ACS/WFC, WFC3/UVIS and WFC3/IR cameras from $\sim10^4$ individual HST visits \citep{Whitmore2016}. The construction of the HSC is achieved by cross-matching sources detected in each individual HST visit \citep{Budavari2012} and calculating the astrometric corrections based on the reference stars from Pan-STARRS, 2MASS, and SDSS and the known distortion patterns of the HST instruments. Currently, the HSC version 2 released in September 2016 includes about 92 million objects with over 300 million measurements obtained from 112 filter/instrument combinations. 

Since some regions of the sky have been observed multiple times by the HST, it provides us a great opportunity to conduct a systematic search for the variable sources at faint magnitudes and high resolution that are inaccessible by the ground-based telescopes. The goal of the Hubble Catalog of Variables (HCV) project funded by the European Space Agency (ESA) is to identify such variable sources based on the HSC by using an automated pipeline containing a set of detection and validation algorithms. The HCV will be released in 2018 at the Mikulski Archive for Space Telescopes (MAST) and the ESA Hubble Science Archives. The HCV pipeline will be deployed at the Space Telescope Science Institute (STScI) so that an updated HCV may be generated following future releases of HSC.

\section{Data Pre-processing and Variability Detection}

Our approach to finding variables is to assume that there is a sea of constant sources with a few variables standing out according to certain criteria. However, the error bars associated with brightness measurements usually do not take into account systematic effects which typically dominate the error budget of bright objects. Thus, the critical point for the variable detection is to set up a correct baseline for the constant sources. This is a challenge for the HCV since the HSC data are inhomogeneous as the HST observations are taken with different filters, instruments, observation strategies and with varying photometric accuracy. Meanwhile, the HSC is constructed based on the visit-combined images so the number of independent measurements is typically far smaller than the number of actual exposures.

To address this issue, several corrections have been made by the HCV pipeline to deliver a clean dataset for the variability detection. During the import of the raw HSC data, the image artifacts and saturated sources are excluded by using the HSC internal flags. Additional cut-offs are applied to remove the measurements associated with double-detection, generated by the old version of the source extraction pipeline, or affected by the ghost images and the existence of edge effect. For each object, the imported data in a given filter are collected to form a lightcurve. Instead of using photometric error, a simple but efficient synthetic error ($Err_{syn}$) has been constructed to evaluate each measurement within a lightcurve based on the photometric error ($Err_{Mag_{Aper2}}$), concentration index (CI), offset distance from the cross-matching position (D), and the difference between circular and elliptical aperture magnitude ($Mag_{Aper2}-Mag_{Auto}$), 
\begin{displaymath}
Err_{syn} = \sqrt{\left(\frac{Err_{Mag_{Aper2}}}{<Err_{Mag_{Aper2}}>}\right)^2 + \left(\frac{CI}{<CI>}\right)^2 + \left(\frac{D}{<D>}\right)^2 + \left(\frac{Mag_{Aper2}-Mag_{Auto}}{<Mag_{Aper2}-Mag_{Auto}>}\right)^2}, 
\end{displaymath}
where the $<X>$ indicates the median value of each parameter. The $Err_{syn}$ is sensitive to several image deficiencies such as cosmic rays, image misalignment and distortion, blending and so on. A robust regression linear function is fitted for each lightcurve to correctly estimate the tendency when taking into account the synthetic error \citep{Press1992}. After calculating the robust sigma of the linear fitting residual ($\sigma$) and the synthetic error residual ($\sigma_{syn}$, based on the median value of $Err_{syn}$), both strong (largely deviating from the linear fit, e.g., $\sigma>3$) and potential outliers (fitted by the linear function but deviating from the median value of $Err_{syn}$, e.g., $\sigma<3$ and $\sigma_{syn}>4$) are identified. Visits having a large percentage (e.g., $>20\%$) of strong outliers are identified as bad images, which are usually caused by image misalignment, and all the associated measurements are discarded. Then a local zero-point correction is calculated for each visit based on the median difference between the predicted magnitudes by the linear function and the observed normal magnitudes (e.g., $\sigma<3$ and $\sigma_{syn}<4$) for all the targets within a radius of 20" around the target being corrected. It should be able to correct various systematic effects, such as variation of PSF, zero-point drifting and the telescope ``breathing''. Finally, all the strong and potential outliers are discarded to reduce the false detection rate of candidate variables. Figure~\ref{sample} shows two examples of lightcurve pre-processing.

\articlefiguretwo{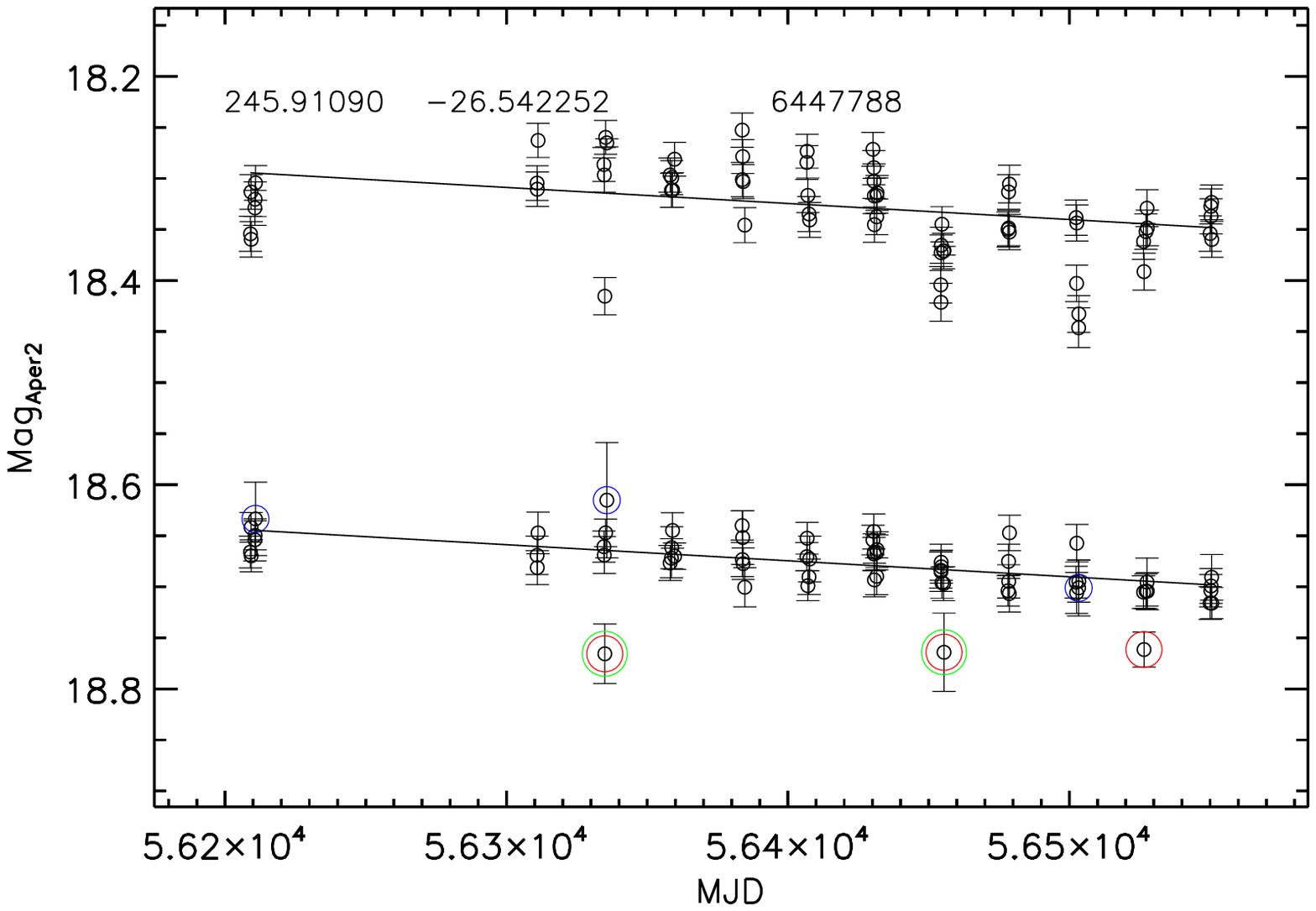}{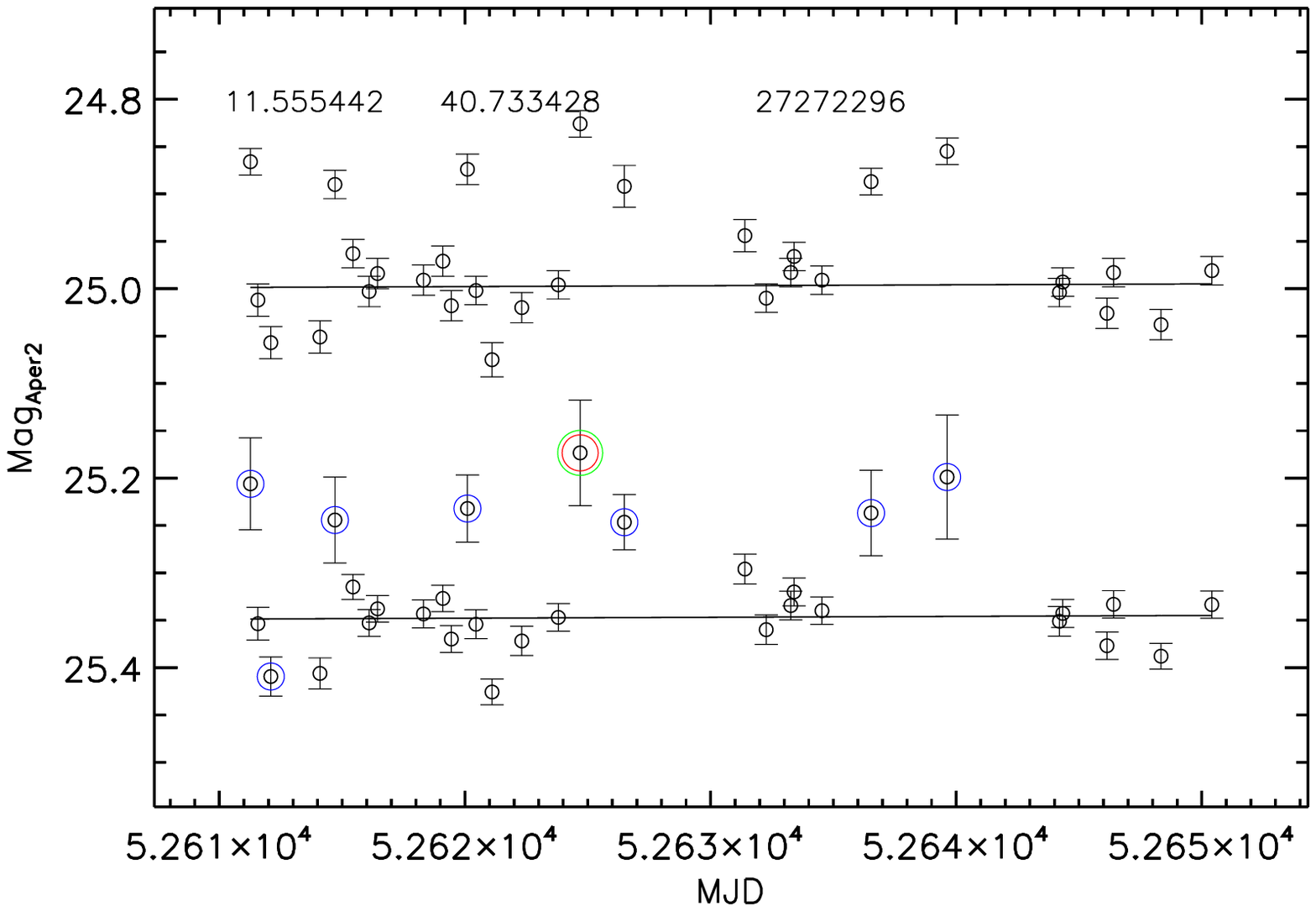}{sample}{Examples of lightcurve pre-processing. For clarity, an offset of 0.35 mag has been added between the raw (top) and pre-processed (bottom) lightcurves. The strong and potential outliers are indicated by red and blue circles, respectively. The measurements that satisfied both criteria of strong and potential outlier are marked by green circles.}

As the dataset is cleaned up, 20 Variability Indexes (VIs) which characterize a wide range of variability features for a lightcurve are computed as listed in Table~\ref{vitable} \citep{Gavras2017, Sokolovsky2017}. The VIs characterize scatter of measurements in the lightcurve and/or its smoothness in various ways. Some VIs take into account the estimated photometric errors, order in which measurements were taken and/or time intervals between measurements.

\begin{table}
  \begin{center}
  \caption{The list of VIs computed by the HCV pipeline.}
  \label{vitable}
 {\footnotesize
  \begin{tabular}{lcccc}
  \\ \hline 
{\bf Index} & {\bf Errors} & {\bf Order} & {\bf Time} 
  \\ \hline
\multicolumn{4}{c}{Scatter-based indexes}  \\
reduced $\chi^2$ statistic - $\chi^2_{red}$  & $\surd$ \\
weighted standard deviation - $\sigma_w$ &$\surd$  \\
median absolute deviation - MAD  \\
interquartile range - IQR  \\
robust median statistic - RoMS& $\surd$ \\ 
normalized excess variance - $\sigma^2_{NXS}$ & $\surd$ \\ 
normalized peak-to-peak amplitude - u& $\surd$ \\
\multicolumn{4}{c}{Correlation-based indexes}\\
Stetson's I index& $\surd$ &$\surd$ & $\surd$ \\
Stetson's J index&  $\surd$ &$\surd$ & $\surd$ \\
time-weighted Stetson's J(time) & $\surd$ &$\surd$ & $\surd$ \\
clipped Stetson's J(clipped) & $\surd$ &$\surd$ & $\surd$ \\
Stetson's L index & $\surd$ &$\surd$ & $\surd$ \\
time-weighted Stetson's L(time) & $\surd$ &$\surd$ & $\surd$ \\
clipped Stetson's L(clipped)& $\surd$ &$\surd$ & $\surd$ \\
consecutive same sign deviations - CSSD & &$\surd$ \\
excursions - $E_x$ & $\surd$ &$\surd$ & $\surd$ \\
autocorrelation - $l_1$&&$\surd$\\
inverse von Neumann ratio - 1/$\eta$&&$\surd$\\
excess Abbe value $\mathcal{E}_A$&  &$\surd$ & $\surd$ \\
$S_B$ statistic & $\surd$ &$\surd$  \\ \hline
  \end{tabular}
  }
 \end{center}
\vspace{1mm}
\end{table}

After extensive testing on the HSC data, the objects having the $MAD>5\sigma$ and $\chi^2_{red}>3$ are considered as candidate variables, since this combination is robust to outliers and sensitive to a broad range of variability types. 

Currently, with at least 5 measurements for each lightcurve, there are $\sim$8000 multi-filter candidate variables (MFCV; at least detected in 2 filters) and $\sim47000$ single-filter candidate variables (SFCV) identified among $\sim$2 million pre-processed targets. The MFCV will also be visually inspected to confirm variability.

\section{Summary}
The HCV is a deep, inhomogeneous catalog of variable sources generated by carefully pre-processing data from the HSC and computing a wide variety of VIs to determine the variability status of each target. It will be available at MAST and the ESA Hubble Science Archives in 2018. The data pre-processing and variability detection techniques used for HCV can be applied to any other time-domain survey.

\acknowledgements The Hubble Catalog of Variables is a three-year project funded by the European Space Agency under contract No.4000112940.

\end{document}